\documentclass[]{article}%10pt12pt
\usepackage[bookmarks = true, pdfpagemode = None, pdfstartview = FitH,colorlinks = true, citecolor = black, linkcolor = black, urlcolor = black]{hyperref} %,\\, backref
\usepackage{url}
\usepackage{amsmath}\interdisplaylinepenalty=2500
\usepackage{amssymb}
\usepackage[matrix,frame,arrow]{xy}
\usepackage{amsmath}
\newcommand{\bra}[1]{\left\langle{#1}\right\vert}
\newcommand{\ket}[1]{\left\vert{#1}\right\rangle}
\newcommand{\qw}[1][-1]{\ar @{-} [0,#1]}
\newcommand{\qwx}[1][-1]{\ar @{-} [#1,0]}
\newcommand{\gate}[1]{*{\xy *+<.6em>{#1};p\save+LU;+RU **\dir{-}\restore\save+RU;+RD **\dir{-}\restore\save+RD;+LD **\dir{-}\restore\POS+LD;+LU **\dir{-}\endxy} \qw}
\newcommand{\meter}{\gate{\xy *!<0em,1.1em>h\cir<1.1em>{ur_dr},!U-<0em,.4em>;p+<.5em,.9em> **h\dir{-} \POS <-.6em,.4em> *{},<.6em,-.4em> *{} \endxy}}
\newcommand{\measuretab}[1]{*{\xy *+<.6em>{#1};p\save+LU;+RU **\dir{-}\restore\save+RU;+RD **\dir{-}\restore\save+RD;+LD **\dir{-}\restore\save+LD;+LC-<.5em,0em> **\dir{-} \restore\POS+LU;+LC-<.5em,0em> **\dir{-} \endxy} \qw}
\newcommand{\control}{*-=-{\bullet}}
\newcommand{\ctrl}[1]{\control \qwx[#1] \qw}
\newcommand{\targ}{*{\xy{<0em,0em>*{} \ar @{ - } +<.4em,0em> \ar @{ - } -<.4em,0em> \ar @{ - } +<0em,.4em> \ar @{ - } -<0em,.4em>},*+<.8em>\frm{o}\endxy} \qw}
\newcommand{\qswap}{*=<0em>{\times} \qw}
\newcommand{\push}[1]{*{#1}}
\newcommand{\gategroup}[6]{\POS"#1,#2"."#3,#2"."#1,#4"."#3,#4"!C*+<#5>\frm{#6}}
\newcommand{\lstick}[1]{*!R!<.5em,0em>=<0em>{#1}}
\newcommand{\Qcircuit}{\xymatrix @*=<0em>}
\usepackage{cases}
\usepackage{amsthm}
\theoremstyle{plain}
\newtheorem{Requirement}{\textbf{Quantum Fault Model}}
\newtheorem{theorem}{Theorem}[section]

\theoremstyle{definition}
\newtheorem{definition}[theorem]{Definition}

\newcommand{\qedsymb}{\hfill{\rule{2mm}{2mm}}}

\theoremstyle{remark}

\begin{document}
\title{Fault models for quantum mechanical switching networks}
\author{J.D. Biamonte,\thanks{The authors are with Portland State University, 1900 SW Fourth Avenue, P.O. Box 751,
Portland, Oregon 97201, USA. JDB present address: Oxford University Computing Laboratory, Wolfson Building, Parks Road, Oxford, OX1 3QD, UK.}~ J.S. Allen and M.A. Perkowski}% etc
\date{\today}
\maketitle

\begin{abstract}
The difference between faults and errors is that, unlike faults, errors can be corrected using control codes.  In classical test and verification one develops a test set separating a correct circuit from a circuit
containing any considered fault.  Classical faults are modelled at the logical level by fault models that act
on classical states.  The stuck fault model, thought of as a lead connected to a power
rail or to a ground, is most typically considered.  A classical test set complete for the stuck fault
model propagates both binary basis states, $0$ and $1$, through
all nodes in a network and is known to detect many physical faults.
A classical test set complete for the stuck fault
model allows all circuit nodes to be completely
tested and verifies the function of many gates.  It is natural to ask if one may adapt any of the known classical methods to test quantum circuits.  Of course, classical fault models do not capture
all the logical failures found in quantum circuits.  The first obstacle faced when using methods from classical test is developing a set of realistic quantum-logical fault models.  Developing fault models to abstract the test problem away from the device level motivated our study.  Several results are established.  First, we describe typical modes of failure present in the physical design of quantum circuits.  From this we develop fault models for quantum binary circuits that enable testing at the logical level.  The application of these fault models is shown by adapting the classical test set generation technique known as constructing a fault table to generate quantum test sets.  A test set developed using this method is shown to detect each of the considered faults.
\end{abstract}

\newpage 

\section{Introduction}
Test methods in use today began to be developed in the 1960's anticipating circuit sizes that would make exhaustive test methods intractable.  Classical test theory attempts to determine if a given circuit is or is not functional~\cite{Kautz:afd:61, kautz:tff:67}.  This is accomplished by testing a circuit in order to determine if any of the logical failures modelled by the set of considered fault models are present.  Fault models are typically inspired by physical failures but may also represent abstractions that enable the creation of test sets detecting an abundance of actual physical faults.

Applying established methods of classical circuit test theory to test quantum circuits has attracted interest in recent times~\cite{testchallenge}, with
preliminary results reported
in~\cite{BiamonteAP:04, HayesPB:04, PerkowskiBL:04, PatelHM:04}.
Despite interest, adequate
justification has not been given for any fault model considered. For instance, much current research has assumed that the classical stuck fault model impacts quantum circuits.  The classical stuck
fault model does not capture non-deterministic and non-localized quantum mechanical faults.  It is even possible to develop classical test sets for reversible circuits complete for the stuck fault model,
that never turn on a gate~\cite{AllenBP:05}.

Experimental physicists who build quantum circuits have not experienced much need to
research optimized testing methods due to the size of the currently attainable
qubit count so the quantum test problem remains of theoretical interest today~\cite{AminGIIMVSZ:04,QuantumSearchNMR:98, countingOnNMRJones:98, SteaneLucas:ION:00}. In appendix~\ref{sec:Current Methods Used to Test Quantum Circuits} we review both this currently practiced (exhaustive) approach to the quantum circuit test problem as well as distance measures to compare quantum states. 

A quantum circuit is built out of gates and these are connected using nodes and wires.  If so many as  a single gate, single node or a wire is broken, the quantum circuit is unusable.  This is in contrast to quantum errors corrected by control codes~\cite{ErrorModelandThresholds:97}.  The possibility to construct quantum networks completely free of faults is generally assumed since error correction relies on fault free networks.  It is therefore important to test the inter-connectivity of the network nodes, wires and gates in an attempt to locate faults. This work addresses this problem by justifying several logical fault models dependent on the structure of quantum switching networks. Test sets complete for logical fault models determine if any considered fault is present in the interconnections or the gates of a quantum network.  In practice, test sets are developed to detect all of the most common faults.

Classically, one defines a testability measure as a product of
observability and controllability. The controllability of testing a
circuit corresponds to propagating a specific input test vector
through a network, such that it will map a test vector to a place of
fault. This represents an added challenge in the case of testing quantum
circuits, since inputs may become entangled and faults may occur non-deterministically. Furthermore, depending on the measurement basis chosen, certain faults may not be detectable.  With these limitations in mind and after defining our notation next, in $\S$~\ref{sec:Fault Models}, we present a set of quantum fault models ($\S$~\ref{sec:Pauli Faults} through $\S$~\ref{sec:Measurement Faults}).  The presence of even one of these faults will make a quantum circuit unusable so we show how to develop test sets in $\S$~\ref{sec:example} under the assumption that the fault-tolerant quantum circuit~\cite{ErrorModelandThresholds:97} under test contains at most one of the considered faults.  We further assume that each quantum circuit is executed multiple times and that the output is averaged over, using a majority voting procedure.\footnote{The Chernoff Bound asserts that polynomial many repetitions of
independent samples will converge exponentially fast to the
true mean.  See for instance chapter 4 of the text book by R. Motwani and P. Raghavan, {\em "Randomized Algorithms,"} Cambridge University Press, (1995).}

\section{Background and notation}\label{sec:Fault Models}

Let us define the state $\sqrt{2}\cdot\ket{\pm}=\ket{0}\pm\ket{1}$ and mention that the notational conventions from the textbook by Nielsen and
Chuang~\cite{NielsenC:00} are used.\footnote{Note to readers new to quantum circuits: The background to read this paper appears in
Chapters 1, 3 and 8 of the text book~\cite{NielsenC:00}.
Course lecture notes
accompanying~\cite{NielsenC:00} can be found online~\cite{Oskin:04}.}  We let $\hat Z$ be the standard observable in the computational basis and $\hat X$ be an observable in the conjugate basis~\cite{DWLeungPhD:00}.\footnote{The conjugate basis is formed by measuring $\hat Z$ after rotating all qubits with a Hardamard transform~\cite{NielsenC:00}.}  We present a set of physically motivated quantum fault models for quantum circuits
built from $k-$CN gates~\cite{NielsenC:00}. To separate the general from the particular, a focus is made on
error and fault models that are independent of implementation.  Without loss of generality
examples from several current technologies such as liquid state nuclear magnetic resonance
spectroscopy (NMR)\footnote{See Jones~\cite{NMRIntroJones:99} for an
introduction geared towards those new to NMR Quantum Computing.} and
optical quantum computation are used. Although the NMR implementation is not scalable, it is
currently the most successful~\cite{fiveBitNMR:01}. Optical quantum computation has the potential of scalability and has recently made progress. For instance, in
$2003$ O'Brien \emph{et al.} successfully demonstrated an optical
quantum CN gate~\cite{O'Brien:CN_Demo:03}.  This gate was improved
and further characterized in $2004$ by both O'Brien \emph{et
al.}~\cite{Obrien:QPT_CNOT:04} and White \emph{et
al.}~\cite{White:CN_Measure:03}.

\subsection{Testing Definitions} 
\begin{figure}[h]\small{
\centerline{
    \Qcircuit @C=1em @R=.2em {
 &                  &                & \mbox{$_1$}    &             &                &             & \mbox{$_2$}   &             &                & \mbox{$_3$} & & \mbox{$_4$} &\\\\\\
 & \lstick{\ket{i_0}} & \qw            & \qswap         & \qw         & \ctrl{5}       & \qw         & \qswap        & \qw         & \ctrl{5}       & \qswap & \ctrl{9} &  \qswap & \qw & \measuretab{m_0}\\\\\\\\
 &                  & \mbox{$_5$}    &                & \mbox{$_6$} &                & \mbox{$_7$} &               & \mbox{$_8$} &                & \mbox{$_9$} & & \mbox{$_{10}$} &\\
 & \lstick{\ket{i_1}} & \qswap         & \ctrl{4}       & \qswap      & \targ          & \qswap      & \ctrl{4}      & \qswap      & \targ          & \qswap & \qw      &  \qswap & \qw& \measuretab{m_1}\\\\\\
 &                  & \mbox{$_{11}$} &                &             & \mbox{$_{12}$} &             &               &             & \mbox{$_{13}$} & & & \mbox{$_{14}$} &\\
 & \lstick{\ket{i_2}} & \qswap         & \gate{\vee}       & \qw         & \qswap         & \qw         & \gate{\vee^\dag} & \qw         & \qswap  &
\qw    & \gate{\vee} &  \qswap & \qw & \measuretab{m_2}}}}
 \caption{\label{tofgate}
 $2-$CN gate with 14 gate external error locations numbered above an $"\times"$ and five possibly erroneous gates.
 The construction of the gates used in this circuit are outlined in Sec.~\ref{subsec:Control Faults}.}
\end{figure}
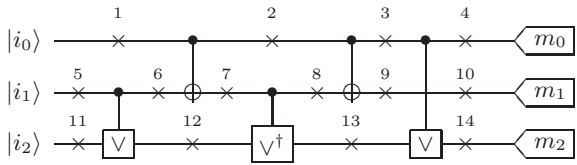

In quantum error correcting codes, fault locations are between
circuit stages,\footnote{In particular
see~\cite{ErrorModelandThresholds:97} where
Knill, Laflamme, and Zurek justified the idea of an error location
for the purpose of quantum error correction.} and have quantifiable
error probabilities~\cite{ErrorModelandThresholds:97}. For example, consider
the five stage circuit shown in Fig.~\ref{tofgate}. The numbered
locations of possible gate external faults are illustrated by
placing an "$\times$" on the lines representing qubits.   The five gates, initial states ($\ket{\psi_0},
\ket{\psi_1}, \ket{\psi_2}$) and measurements ($M_0, M_1, M_2$) may also
contain errors.

\begin{definition}\label{Error Location}
    Error Location: The wire locations between stages as well as any node or gate in a given network  represent error locations.
\end{definition}

\begin{definition}\label{def:Gold Circuit}
    Gold Circuit: An ideal quantum circuit denoted $GC$.
    Typically a non-ideal quantum circuit is denoted $QC$.
\end{definition}

\begin{definition}\label{def:Complete Fault Set}
    Fault Set: Denote by $F_q$ a set containing all considered faults assumed to impact
    $QC$.
\end{definition}

\begin{definition}\label{def:Quantum Test Pattern}
    Quantum Test Set: A sequence of initial
    states $\ket{\psi_i}$ and corresponding measurements $M_i$ used to distinguish $QC$
    possibly perturbed by any $f \in F_q$ from a gold circuit
    $GC$.
\end{definition}

Complete fault coverage occurs if executing a test set can determine that
all of the considered faults are not present in a given
circuit.

\begin{definition}\label{Fault Coverage}
    Fault Coverage: Denote by $QC$ a quantum circuit possibly perturbed by any element of a set of faults
    $f\in F_q$ and a test set $T$ complete for all $f\in F_q$.  Fault Coverage
    occurs for fault $f$ by experimentally running $t\subseteq T$ that
    detects $f$.  A quantum test set that detects all considered faults is
    a complete test set.
\end{definition}

Since even the presence of one of the faults considered in this work would make a quantum circuit unusable, when developing quantum test sets in $\S$~9, we assume that the fault-tolerant quantum circuit~\cite{ErrorModelandThresholds:97} under test contains at most one of the considered faults.   This assumption is called the \emph{quantum
single fault model}. Classically, test sets complete for single faults are known to detect the presence of multiple faults~\cite{McCluskey:SFvsD:00}.  Whether or not the quantum single fault model dominates multiple quantum faults is left as an open problem.

\begin{definition}\label{def:Single Quantum Fault Model}
    Quantum Single Fault Model: Consider the quantum fault set $F_q$. In the quantum single fault model,
    test plans are developed to detect all $f\in F_q$ assuming that at most one fault, $f$, is present in $QC$.
\end{definition}

\begin{definition}\label{def:paulimatrix}
Pauli Matrices:
\begin{gather}\label{eqn:sigma}
\sigma_{x}=\ket{1}\bra{0}+\ket{0}\bra{1},\qquad
 \sigma_{y}=i\ket{0}\bra{1}-i\ket{1}\bra{0},\\\nonumber
\sigma_{z}=\ket{0}\bra{0}-\ket{1}\bra{1}\qquad \text{and}\qquad\sigma_{i} =
\ket{0}\bra{0}+\ket{1}\bra{1}.
\end{gather}
\end{definition}

\begin{definition}\label{def:rotationmats}
Rotation Matrices:
\begin{gather}\label{eqn:r}
R_{x}(\theta)=\left(%
 \begin{array}{cc}
  \cos\left(\theta/2\right) & -i \cdot \sin\left(\theta/2\right) \\
  -i \cdot \sin\left(\theta/2\right) & \cos\left(\theta/2\right) \\
\end{array}%
\right),\qquad
R_{y}(\theta)=\left(%
 \begin{array}{cc}
  \cos\left(\theta/2\right) & -\sin\left(\theta/2\right) \\
  \sin\left(\theta/2\right) & \cos\left(\theta/2\right) \\
\end{array}%
\right)\\\nonumber \qquad\text{and}\qquad
R_{z}(\phi)= e^{-i \phi/2}\ket{0}\bra{0}+e^{i \phi/2}\ket{1}\bra{1}.
\end{gather}
\end{definition}

With the basic definitions behind us, the coming sections discuss
gate level failures and introduce several requirements that a quantum test set must satisfy. Each requirement must be satisfied for a test set to be complete and cover all considered
faults.

%For molecules with several coupled spins, the sequence of Figure 1
%must be expanded with extra pulses to refocus undesired J -coupling

\section{Faults modelled with Pauli matrices}\label{sec:Pauli Faults}

\begin{definition}\label{def:Pauli Single Fault Model}
    Pauli Fault Model: The addition of an
    unwanted Pauli matrix $f$ in quantum network $QC$, at error location $l$ and with placement probability $p$.
\end{definition}

The unavoidable entanglement between a quantum processor and the outside
world is described by many authors as, ``\emph{coupling to an
initially independent environment}~\cite{IntroQErrorCorrection:02}.'' This is a primary source of decoherence.  A large amount of research has
been devoted to removing the local effects of decoherence by quantum error
correcting codes. Consequently, what are known as error models are found in the
quantum error correcting code
literature~\cite{shorFaultTol:97, Aharonov:00, Bettelli:04, Barenco:98, oraclenoise}.
The most investigated error model is the ``\emph{independent
depolarizing error}~\cite{ErrorModelandThresholds:97}.'' This model
has the effect of completely randomizing a given qubit with some
probability~\cite{NielsenC:00} and ``\emph{...error models designed to
control depolarizing errors apply to all independent error
models}~\cite{IntroQErrorCorrection:02}.'' These codes are designed
to correct unwanted single qubit $\sigma_x$, $\sigma_y$ and $\sigma_z$ rotations. The following is a list of the range of errors modeled assuming
Pauli Faults, with some supporting references:
\begin{center}
\begin{itemize}\label{list:faults}
    \item Depolarizing Channels~\cite{NielsenC:00, IntroQErrorCorrection:02}\vspace{-.1in}
    \item Amplitude Dampening~\cite{NielsenC:00, ErrorModelandThresholds:97}\vspace{-.1in}
    \item Phase Damping~\cite{NielsenC:00, Kak:99}\vspace{-.1in}
    \item Phase-Flips~\cite{NielsenC:00, ErrorModelandThresholds:97, Kak:99}\vspace{-.1in}
    \item Bit-Flips~\cite{NielsenC:00, ErrorModelandThresholds:97}\vspace{-.1in}
    \item Initialization Inaccuracies~\cite{Kak:99, PerparingHighPurityState}\vspace{-.1in}
    \item Measurement
    Inaccuracies~\cite{qubitMeasure:01, percisionmeasure, ClearwaterBook:98}
\end{itemize}
\end{center}

In addition to noise, repeatable errors in a physical construction
lead to another class of faults addressed in the literature.  These are known as \emph{systematic errors} and are again modeled by Pauli Faults. Systematic errors are closer to the
types of errors that classical test engineers refer to as
faults.  These errors are described by Cummins and Jones
as, ``\emph{arising from the reproducible imperfections in the
apparatus used to implement quantum
computations}~\cite{CumminsJones:00}.'' The most common systematic errors are given in the following list, again with supporting references:

\begin{center}
\begin{itemize}\label{list:faults}
    \item Pulse Length Errors~\cite{measureNMRRotorFidelity:01,CumminsJones:00,TacklingSystematicErrors,ObenlandD:96}\vspace{-.1in}
    \item Off-Resonance Effects~\cite{CumminsJones:00, measureNMRRotorFidelity:01, Vandersypen:NMRBasics:02}\vspace{-.1in}
   \item Refocusing Errors~\cite{rofocussingNMR,Vandersypen:NMRBasics:02}
   \end{itemize}
\end{center}

\begin{Requirement}\label{Axiom:bit flip}
    (\textbf{Pauli Fault Model 1}) A bit flip $(\sigma_x$ or $\sigma_y)$ at any error location must be detectable.~$\blacksquare$
\end{Requirement}

It turns out that these faults are very easy to detect in quantum switching circuits.  The following Theorem~(\ref{lemma:Pauli Faults compbasis easy to detect}) states that any $\sigma_x$ or $\sigma_y$ fault occurring in a network built from $k-$CN gates is detectable with any computational basis input
state. In addition, a reversible system preserves information and one can show that the probability of detection
for fault $f$ observable with $\hat{A}$, is directly related
to the probability of $f$'s presence.
\begin{theorem}\label{lemma:Pauli Faults compbasis easy to detect}
    Pauli Faults $\sigma_x$ and $\sigma_y$ impacting an $n$ qubit network $QC$
    comprised of $k-$CN gates at any gate external error location are detected with any basis     input state $\ket{s}$ given an observable in the computational basis.
\end{theorem}
\begin{proof}
    Any reversible binary quantum network $QC$ of $l$ qubits and $n$ stages bijectively maps each input $\ket{s}$ to a unique output $\ket{s'}$.
    Each gate $g\in QC$ is reversible, thus any stage $n$ acting on input vector $\ket{s}$ corresponds to
    exactly one output vector at the $(n+1)^{th}$ stage.  A qubit flip ($f_p$) occurs before or
    after any stage $n$ on any wire $l$, changing the output from the previous
    stage (and input to the next).  Any $\sigma_x$ or $\sigma_y$ fault is therefore detectable based on the properties
    of reversibility~\cite{Zurek:RSIPS:84} with any basis state input $\ket{k}$ given an observable
    in the computational basis.
\end{proof}
From Theorem~\ref{lemma:Pauli Faults compbasis easy to detect} it follows that any test set that contains a basis state input and corresponding measurement in the computational basis dominates a test set complete for Pauli Fault Model~\ref{Axiom:bit flip}.  This desirable property does not hold for $\sigma_z$ phase faults.

\begin{Requirement}\label{Axiom:phase flip}
    (\textbf{Pauli Fault Model 2}) A phase flip $(\sigma_z)$ at any error location must be detectable.~$\blacksquare$
\end{Requirement}

\section{Initialization Faults}\label{subsec:Initialization Errors}
Initialization faults were discussed in detail by Kak~\cite{Kak:99},
and addressed experimentally
in~\cite{PerparingHighPurityState, Vandersypen:NMRBasics:02}.  Because initialization
accuracies relate to a machine's ability to perform a task (such as
not altering the initial state population in
NMR~\cite{Vandersypen:NMRBasics:02}), one can develop a test set to determine if the machine is impacted by any of the considered initialization faults. We model these
faults using Def.~\ref{def:Single Rotation
Initialization Fault}.

\begin{definition}\label{def:Single Rotation Initialization Fault}
    Initialization Error: A qubit with an initial state impacted by an unwanted
    rotation $R_n(\theta)$ where $n\in\{x, y, z\}$, \emph{or} a qubit that
    is only correctly prepared in one basis state and not the other.
    %(Modeled by acting on qubit $m$ with one of Eqns..~\ref{eqn:rx},~\ref{eqn:ry} or~\ref{eqn:rz}).
\end{definition}

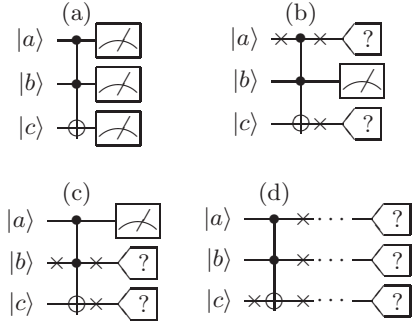
\begin{figure}[t]
\small{\centerline{ $\begin{array}{cc} \Qcircuit @C=.4em @R=.4em {
  & \mbox{~(a) } & &&&&&&&\\
\lstick{\ket{a}} & \ctrl{1} & \meter\\
\lstick{\ket{b}} & \ctrl{1} & \meter \\
\lstick{\ket{c}} & \targ & \meter } & \Qcircuit @C=.4em @R=.4em {
  & &\mbox{~(b) } & \\
\lstick{\ket{a}} & \qswap & \ctrl{1} & \qswap & \qw & \measuretab{?}\\
\lstick{\ket{b}} & \qw    & \ctrl{1} & \qw & \qw & \meter\\
\lstick{\ket{c}} & \qw    & \targ & \qswap & \qw & \measuretab{?}} \\\\
  \Qcircuit @C=.4em @R=.4em {
  & &\mbox{~(c) } & &&&&&&&&\\
\lstick{\ket{a}} & \qw    & \ctrl{1} & \qw & \qw & \meter\\
\lstick{\ket{b}} & \qswap & \ctrl{1} &  \qswap& \qw& \measuretab{?}\\
\lstick{\ket{c}} & \qw    & \targ &  \qswap & \qw& \measuretab{?}} &
\Qcircuit @C=.4em @R=.4em {
  & &\mbox{~(d) } & \\
\lstick{\ket{a}} & \qw    & \ctrl{1} & \qw & \qswap &  \qw && \mbox{$\cdots$}&&& \qw &\measuretab{?}\\
\lstick{\ket{b}} & \qw    & \ctrl{1} & \qw & \qswap & \qw && \mbox{$\cdots$}&&& \qw &\measuretab{?}\\
\lstick{\ket{c}} & \qswap & \targ & \qw & \qswap & \qw && \mbox{$\cdots$}&&& \qw &\measuretab{?}} \\
\end{array}$}}
\caption{Initialization Errors Impacting a $2-$CN Gate: (a) correct
circuit, (b)-(d) various initialization errors.}\label{initCN}
\end{figure}

From Def.~\ref{def:Single Rotation Initialization Fault}, examples of how initialization faults spread are shown in
Fig.~\ref{initCN}. The
fault in Fig.~\ref{initCN}~(b) could occur when
the desired initial state is $\ket{01c}$ and the top qubit is
inverted ($\ket{c}\rightarrow
\ket{\bar{c}}$) resulting in the state:
$\cos\theta\ket{01c}-i\sin\theta\ket{11\bar{c}}$. After being acted on by the $2-$CN gate, the state of the system becomes
$\cos\theta\ket{01c}-i\sin\theta\ket{11c}$. There is now a
probability of $(\sin\theta)^2$ that an incorrect value will be
measured. %This uncertainty is denoted by replacing \Qcircuit @C=.1em
%@R=.1em { & \qw\qw & \qw & \qw & \qw & \meter &&&&\\&&} symbols with
%\Qcircuit @C=.1em @R=.04em { & \qw\qw & \qw & \qw & \qw &
%\measuretab{?} &\\&}.

A similar scenario holds for Fig.~\ref{initCN}~(c)---in this case
the center qubit is impacted by an initialization fault as
opposed to the top qubit in Fig.~\ref{initCN}~(b).  In
Fig.~\ref{initCN}~(d) the desired initial state is $\ket{--+}$, however a fault impacts the bottom qubit
flipping its phase---changing the initial state to $\ket{---}$.  Now the $2-$CN gate will entangle the
state of the system incorrectly, resulting in state
$(\ket{00}-\ket{01}-\ket{10}-\ket{11})\ket{-}$. This fault can be
detected with a test set detecting unwanted instances of the Pauli
Faults. A second
type of initialization error will now be discussed.

Generally, there exists a certain set of states left invariant under
a quantum operation~\cite{NielsenC:00}.  For example, qubit
preparation may be altered with a form of amplitude dampening.
Thus, a faulty qubit might only allow preparation into one state. A qubit that can only be prepared in density state $\ket{0}\bra{0}$ is
modeled with the following operation elements:
\begin{equation*}
E_0 = \ket{0}\bra{0}+\left(\sqrt{1-\gamma}\right)\ket{1}\bra{1}
%\left[
%         \begin{array}{cc}
%           1 & 0 \\
%           0 & \sqrt{1-\gamma} \\
%         \end{array}
%       \right]
\end{equation*}
\begin{equation*}
E_1 = \ket{0}\bra{0}+\left(\sqrt{\gamma}\right)\ket{1}\bra{1}
%\left[
%         \begin{array}{cc}
%           0 & \sqrt{\gamma} \\
%           0 & 0 \\
%         \end{array}
%       \right]
\end{equation*}

Consider a register prepared in density state $\rho =\rho_0
\otimes \ldots \otimes \rho_k \otimes \ldots \otimes\rho_n$. The
$k^{th}$ qubit is desired to start the computation in an arbitrary
state, expressed as:
\begin{equation*}
    \rho = \rho_0
\otimes \ldots \otimes\left(
                \begin{array}{cc}
                 \alpha^2 & \alpha\beta^* \\
                 \beta\alpha^* & \beta^2 \\
               \end{array}
             \right)\otimes \ldots \otimes\rho_n
\end{equation*}
The unwanted impact of initial state dampening is expressed as:
$\mathcal{E}(\rho)=\sum_k \sigma_i \otimes \ldots \otimes E_k\otimes \ldots
\otimes \sigma_i  \cdot \rho \cdot \sigma_i \otimes \ldots \otimes
E_k^\dagger\otimes \ldots \otimes \sigma_i $.  This results in the state:
\begin{equation*}
\rho' =\rho_0 \otimes \ldots \otimes \left(
                \begin{array}{cc}
                 \alpha^2+\gamma\beta^2 & \alpha\beta^*\sqrt{1-\gamma} \\
                 \beta\alpha^2\sqrt{1-\gamma} & \beta^2(1-\gamma) \\
               \end{array}
             \right) \otimes \ldots \otimes\rho_n.
\end{equation*}
The projection of the $k^{th}$ qubit's state onto the basis
$\ket{1}\bra{1}$ is forced into basis state $\ket{0}\bra{0}$ and the
off diagonal terms are suppressed (both based on some parameter
$\gamma$). Similarly, a faulty qubit might only allow preparation into density state
$\ket{1}\bra{1}$. A test set complete for Quantum Fault Model~\ref{Requirement:initialization fault model} determines if any qubit can only be correctly prepared in one basis state and not the other.

\begin{Requirement}\label{Requirement:initialization fault model}
    (\textbf{Initialization Fault Model}) Each qubit must be initialized in both basis states $\ket{0}$ and $\ket{1}$.~$\blacksquare$
\end{Requirement}

\section{Lost Phase Faults}\label{subsec:Phase Faults}
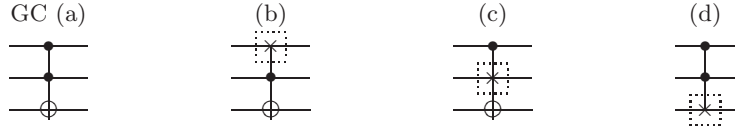
\begin{figure}[h]\small{
\centerline{ $\begin{array}{cccc}
  \Qcircuit @C=1.2em @R=.5em @!R {
  & \mbox{~GC~(a) } &&&&& \\
%&\ctrl{1} & \qw \\
&\ctrl{1} & \qw\\
&\ctrl{1} & \qw\\
&\targ    & \qw}
 & \Qcircuit @C=1.2em @R=.5em @!R {
  & \mbox{~(b) } &&&&& \\
&\qswap & \qw \gategroup{2}{2}{2}{2}{1.2em}{..}\\
&\ctrl{1}\qwx & \qw\\
%&\ctrl{1} & \qw\\
&\targ    & \qw}
 & \Qcircuit @C=1.2em @R=.5em @!R {
  & \mbox{~(c) } &&&&& \\
&\ctrl{2} & \qw \\
&\qswap & \qw\gategroup{3}{2}{3}{2}{1.2em}{..}\\
%&\ctrl{1} & \qw\\
&\targ    & \qw}
 & \Qcircuit @C=1.2em @R=1.3em @!R {
  & \mbox{~(d) } & \\
&\ctrl{1} & \qw \\
&\ctrl{1} & \qw\\
%&\qswap & \qw\gategroup{4}{2}{4}{2}{1.2em}{..}\\
&\qswap  & \qw\gategroup{4}{2}{4}{2}{1.2em}{..}} \\
\end{array}$}}
\caption{$2-$CN Gate and Phase Faults:~(a) Gold Circuit, (b)
weak top control, (c) weak second control, (d) weak gate.}\label{cir:missing_control_$2-$CN}
\end{figure}

Shenvi, Brown and Whaley~\cite{oraclenoise} studied Grover's search
algorithm~\cite{NielsenC:00} impacted by random phase errors in the
oracle.\footnote{Grover's original algorithm has recently been updated by
L. Grover~\cite{GroverNew:05}. Although this new ``\emph{Fixed Point
Algorithm}'' is more robust, it is still subject to phase errors.  The algorithm
has been experimentally verified by Xiao
and Jones~\cite{Jones:GroverFixedPoint:05}, systematic errors
in the physical implementation were briefly explored. B. Reichardt
and L. Grover have also recently developed methods of systematic
error correction for this new algorithm~\cite{Reichardt:05}.} They model errors by
applying unwanted phase shifts $\pm \epsilon$ to the state of a
quantum register marked by an oracle: $\mathcal{O}:\ket{k}\longrightarrow e^{i\pi\cdot (f(k) \pm \epsilon)}\ket{k}$.
Several references including~\cite{oraclenoise} call this a
"\emph{phase-kick-error}." Fig.~\ref{cir:missing_control_$2-$CN} (b), (c) and (d) illustrate faulty controls that have phase kick-back faults.  A correct $3-$CN gate will map an input state $\ket{+++}\ket{-}$
to output state
$(\ket{000}+\ket{001}+\ket{010}+\ket{011}+\ket{100}+\ket{101}+\ket{110}-\ket{111})\ket{-}$.
Each term in the
superposition that activates the gate will undergo a phase
shift of $\ket{n}\longrightarrow e^{i\pi}\ket{n}$. If the fault in
Fig.~\ref{cir:missing_control_$2-$CN}~(b) is present, the circuit's
output is
$(\ket{000}+\ket{001}+\ket{010}-\ket{011}+\ket{100}+\ket{101}+\ket{110}-\ket{111})\ket{-}$.
In this case, relative phase shifts occur on both states $\ket{011}$
and $\ket{111}$ since those activate the gate when the top
control is broken.  Another type of fault is phase damping.  This is a noise process altering relative phases between quantum states~\cite{NielsenC:00}.  A test set complete for the Lost Phase Fault Model determines if the faults described above are present in a given quantum circuit.
%if any gates in the network applies unwanted relative phase shifts or
%unwanted phase swaps (between target and control) to a quantum state or if any

\begin{Requirement}\label{Requirement:phase1}
    (\textbf{Lost Phase Fault Model}) Consider the circuit input as $H^{\otimes n}\cdot \ket{x_1,x_2,...,x_n}$ with $x_i\in\{0,1\}$ and measurement of $\hat X$.  Given freedom in the choice of input vector $x_1,x_2,...,x_n$ a circuit must be shown to have no missing single control or single gate.~$\blacksquare$
\end{Requirement}

\begin{table}
\centerline{\small{
\begin{tabular}{c||c|c|c|c|c}
     term     & initial& GC (a)  & (b) & (c) & (d) \\
     \hline
  $\ket{000}$ & +1     & +1    &  +1   & +1   & +1 \\
  $\ket{001}$ & $-1$   & $-1$  &  $-1$ & $-1$ & $-1$ \\
  $\ket{010}$ & +1     & +1    &  $-1$ & +1   & +1 \\
  $\ket{011}$ & $-1$   & $-1$  &  +1   & $-1$ & $-1$ \\
  $\ket{100}$ & +1     & +1    &  +1   & $-1$ & +1 \\
  $\ket{101}$ & $-1$   & $-1$  &  $-1$ & +1   & $-1$\\
  $\ket{110}$ & +1     & $-1$  &  $-1$ & $-1$ & +1 \\
  $\ket{111}$ & $-1$   & +1    &  +1   & +1   & $-1$\\
\end{tabular}}}\caption{The impact Phase Faults from Fig~\ref{cir:missing_control_$2-$CN} have on input state $\ket{++-}$.  The first column shows the phase of each term before being acted on by the circuit.  The column GC (a) shows the correct phase relative phase of each term in the superposition.  The remaining columns (b-d) show how the phase changes depending on the fault present.}
\end{table}

To show that the considered phase faults are not present in the network we place Hardamard gates at the start and at the end of the circuit under test and measure $\hat Z$.  Let us denote the circuit under test as $C_i$ where subscript $i$ signifies that impact of the $i^{th}$ considered fault.  We will define the input as $\rho = \ket{x_1x_2x_3}\bra{x_1x_2x_3}$ with $x_1,x_2$ and $x_3$ in $\{0,1\}$. The goal is to show separation of GC from each of the considered faults where $\rho' = H^{\otimes3}\cdot C_i\cdot H^{\otimes3}\cdot\rho\cdot H^{\otimes3}\cdot C_i^\dagger\cdot H^{\otimes3}$ allows freedom in the choice of the input $\rho$.  Here we choose $\rho = \ket{001}\bra{001}$ and note that for fault (b): $\text{tr}(\ket{100}\bra{100}\cdot\rho')=1$, for fault (c): $\text{tr}(\ket{101}\bra{101}\cdot\rho')=1$ and for fault (d): $\text{tr}(\ket{001}\bra{001}\cdot\rho')=1$.  For GC $\text{tr}(\ket{001}\bra{001}\cdot\rho')=\text{tr}(\ket{100}\bra{100}\cdot\rho')=\text{tr}(\ket{101}\bra{101}\cdot\rho')=\text{tr}(\ket{111}\bra{111}\cdot\rho')=1/4$.
To test the considered $2-$CN gate for the considered phase faults one must show separation in the above quantities.

%\begin{Requirement}\label{Requirement:lost phase}
%    (\textbf{Lost Phase Fault Model 2}) Relative Dephasing: Provided the state of the target is $\ket{+}$: relative phase must be shown not to change under arbitrary activating
%    state $\ket{a}$.
%    Furthermore, relative phase must not change under arbitrary non-activating
%    state $\ket{n}$.~$\blacksquare$
%\end{Requirement}

\section{Faded Control Faults}\label{subsec:Control Faults}

The building blocks needed to implement any quantum algorithm with
NMR can be based on single spin rotations and CN
gates~\cite{Vandersypen:NMRBasics:02}.  CN gates are realized using
a scheme illustrated in Fig.~\ref{cir:CN made from H CZ
H}.  The center gate is a CZ gate and is built using a $\phi$
gate with angle $\pi$ (see~\cite{Jones:QLGNMR:98}, $\S~3.1$
Eqn.~$34$):
\begin{equation}\label{eqn:cz_ideal}
    CZ_i = \ket{00}\bra{00} + \ket{01}\bra{01} + \ket{10}\bra{10} +
    e^{i\pi}\ket{11}\bra{11}.
\end{equation}

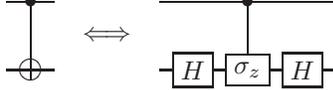
\begin{figure}[h]\centerline{
\Qcircuit @C=.5em @R=0em @!R {
& \ctrl{2} & \qw & & & \qw & \ctrl{2} & \qw & \qw\\
                 & & & \push{\rule{.3em}{0em}\Longleftrightarrow\rule{.3em}{0em}} & & & & & & & \\
& \targ    & \qw & & & \gate{H} & \gate{\sigma_z} & \gate{H} & \qw }
}\caption{CN gate constructed with elementary building
blocks~\cite{Jones:QLGNMR:98}. The H gate ($H=iR_y(\pi/2)R_z(\pi)$)
is given as
$H=\frac{1}{\sqrt{2}}(\ket{0}+\ket{1})\bra{0}+\frac{1}{\sqrt{2}}(\ket{0}-\ket{1})\bra{1}$
and the center controlled phase shift gate (CZ$_i$) is given in
Eqn.~\ref{eqn:cz_ideal}.}\label{cir:CN made from H CZ H}
\end{figure}

In any physical implementation, the $CZ_i$ gate might deviate according to our ability to
apply phase $e^{i\phi}$ correctly to term $\ket{11}\bra{11}$.  This
can be represented as:
\begin{equation}\label{eqn:cz_real}
    CZ_r = \ket{00}\bra{00} + \ket{01}\bra{01} + \ket{10}\bra{10} + e^{i\phi}
    \ket{11}\bra{11}.
\end{equation}
An ideal CN gate creates the
following mapping: $CN_i: \ket{10} \longrightarrow \ket{11}$. If the CZ gate applies a phase at the wrong
angle $\phi$, the mapping becomes: $CN_r: \ket{10} \longrightarrow
(1 + e^{i\phi})\ket{10} + (1 - e^{i\phi})\ket{11}$. The fidelity\footnote{See Appendix~\ref{sec:Current Methods Used to Test Quantum Circuits} for an explanation of the fidelity.}
between the real and ideal CN gate with input $\ket{10}$ is:
\begin{equation*}
    F(CN_i\ket{10}\bra{10}CN_i^\dagger, CN_r\ket{10}\bra{10}CN_r^\dagger) = \frac{1}{2}(1 - \cos\phi).
\end{equation*}
Another gate constructed using
a $\phi$ gate with angle $\pi/2$~\cite{Jones:QLGNMR:98} is known as
the CV gate. The V gate is given as:
\begin{equation}\label{eqn:V Gate}
\vee = \ket{\vee_0}\bra{0} + \ket{\vee_1}\bra{1}, %\footnote{$\vee$ is
%sometimes called the square-root-of-not gate.  The squared magnitude
%of $\vee$ is written as $|\vee |^2 = \vee^\dagger \cdot \vee$.
%Taking the conjugate transpose $\wedge \cdot \vee$, and combining
%overlap via the dot product leads directly to the $N$ gate.}
\end{equation}
where $\ket{\vee_0}=(1+i)\ket{0}+(1-i)\ket{1}$ and
$\ket{\vee_1}=(1-i)\ket{0}+(1+i)\ket{1}$.  The CN and CV gates may
be combined to create $2-$CN gates as shown in Fig.~\ref{cir:$2-$CN
made from CV CN}.  It turns out that by adjusting $\phi$, $n^{th}$
root of NOT gates can be constructed~\cite{ElementaryGates:95}.
These can be used to build any $k-$CN gate (for instance, by setting $\phi=\pm
\pi/4$ the $4^{th}$ root of NOT gates can be created and used to
build the $3-$CN gates in this paper).

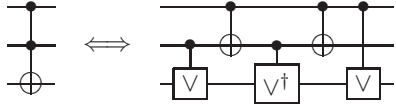
\begin{figure}[h]\centerline{
\Qcircuit @C=.5em @R=0em @!R {
& \ctrl{1} & \qw & & & \qw & \ctrl{1} & \qw & \ctrl{1} & \ctrl{2} & \qw\\
& \ctrl{1} & \qw & \push{\rule{.3em}{0em}\Longleftrightarrow\rule{.3em}{0em}} & & \ctrl{1} & \targ & \ctrl{1} & \targ & \qw & \qw\\
& \targ    & \qw & & & \gate{\vee} & \qw & \gate{\vee^\dag} & \qw &
\gate{\vee} & \qw }} \caption{$2-$CN gate constructed with
elementary building blocks.}\label{cir:$2-$CN made from CV CN}
\end{figure}

Test sets complete for the faded control fault model introduced below must turn each gate on by concurrently
activating all controls~\cite{AllenBP:05}.  These test sets also determine if each control can be turned off properly. A test set complete for the Faded Control Fault Model tests
a controls' function with the target in a
basis state.  It also tests the controls' impact on both activating
and non-activating states.

\begin{Requirement}\label{Requirement:fadedcontrol}
    (\textbf{Faded Control Fault Model}) For the target acting on basis state $\ket{0}$ or $\ket{1}$:
    All controls must be activated concurrently and each
    control must be addressed with a non-activating state.~$\blacksquare$
\end{Requirement}

%\begin{figure}[h]\centerline{
%\small{ $\begin{tabular}{|c|c|c|c|c||c|c|c|c|c|}
%  \hline
%  % after \\: \hline or \cline{col1-col2} \cline{col3-col4} ...
%  \emph{input} & GC(a) & b & c & d & \emph{input} & GC(a) & b & c & d \\
%  \hline
%  $\ket{0}$ & 0 & 0 & 0 & 0 & $\ket{8}$  & 8 & 8 & 8 & 8\\
%  \hline
%  $\ket{1}$ & 1 & 1 & 1 & 1 & $\ket{9}$ & 9 & 9 & 9 & 9\\
%  \hline
%  $\ket{2}$ & 2 & 2 & 2 & 2 & $\ket{10}$ & 10 & 10 & 10 & 10\\
%  \hline
%  $\ket{3}$ & 3 & 3 & 3 & \textbf{11} & $\ket{11}$ & 11 & 11 & 11 & \textbf{3}\\
%  \hline
%  $\ket{4}$ & 4 & 4 & 4 & 4 & $\ket{12}$ & 12 & 12 & 12 & 12\\
%  \hline
%  $\ket{5}$ & 5 & 5 & \textbf{13} & 5 & $\ket{13}$ & 13 & 13 & \textbf{5} & 13\\
%  \hline
%  $\ket{6}$ & 6 & \textbf{14} & 6 & 6 & $\ket{14}$ & 14 & \textbf{6} & 14 & 14\\
%  \hline
%  $\ket{7}$ & 15 & 15 & 15 & 15 & $\ket{15}$ & 7 & 7 & 7 & 7\\
%  \hline
%\end{tabular}$}
%} \caption{Truth Table containing decimal entries for the a $3-$CN
%Gate Impacted by Faded Control Faults. The column denoted 'input'
%shows the possible input combinations. Columns denoted 'b' to 'd'
%illustrate the circuit's response missing the top, second and bottom control, respectively.}\label{tab:Truth Table for Missing Control Faults}
%\end{figure}

\begin{table}
\centerline{\small{
\begin{tabular}{c||c|c|c}
     input    & GC (a)        & (b) & (c) \\
     \hline
  $\ket{000}$ & $\ket{000}$   & $\ket{000}$  & $\ket{000}$ \\
  $\ket{001}$ & $\ket{001}$   & $\ket{001}$  & $\ket{001}$ \\
  $\ket{010}$ & $\ket{010}$   & $\ket{0\underline11}$  & $\ket{011}$ \\
  $\ket{011}$ & $\ket{011}$   & $\ket{01\underline0}$  & $\ket{010}$ \\
  $\ket{100}$ & $\ket{100}$   & $\ket{100}$  & $\ket{10\underline1}$ \\
  $\ket{101}$ & $\ket{101}$   & $\ket{101}$  & $\ket{10\underline0}$ \\
  $\ket{110}$ & $\ket{111}$   & $\ket{111}$  & $\ket{111}$ \\
  $\ket{111}$ & $\ket{110}$   & $\ket{110}$  & $\ket{110}$ \\
\end{tabular}}}\caption{Fault table for $2-$CN gate with (b) missing top control and (c) missing bottom control.  A test set complete for this fault table is as follows: To separate GC from (b) and (c) measure $\hat Z$ for the following inputs $\ket{01\star}$ and $\ket{10\star}$, with $\star\{0,1\}$.}
\end{table}

\section{Forced Gate Faults}\label{subsec:Broken Gate
Faults}

%Any gate model quantum algorithm may be depicted as a
%sequence of unitary transformations represented as rotations in a
%Hilbert space. In NMR, "\emph{...examples of unitary transformations
%are evolution during RF pulses and free evolution under the system
%Hamiltonian...}~\cite{Vandersypen:NMRBasics:02}." In practice,
%"\emph{non-qubit degrees of freedom are
%possible}~\cite{extended_hilbert:04}" and experimentally quantum
%gates perform a process which approximates the desired unitary
%operation, that adds decoherence~\cite{NielsenC:00}. (In NMR,
%"\emph{...relaxation processes give rise to non-unitary
%transformations}~\cite{Vandersypen:NMRBasics:02}...")

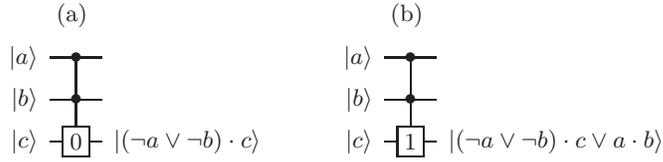
\begin{figure}[h]\small{\centerline{$
\begin{array}{cc}
      \Qcircuit @C=.5em @R=.5em @!R {
  & \mbox{(a) } &&&&&&&&&& \\
\lstick{\ket{a}} & \ctrl{1} & \qw \\
\lstick{\ket{b}} & \ctrl{1} & \qw \\
\lstick{\ket{c}} & \gate{0} & \qw &\push{\ket{(\neg a \vee \neg b)\cdot c}}} &
   \Qcircuit @C=.5em @R=.5em @!R {
  & \mbox{(b) } & \\
\lstick{\ket{a}} & \ctrl{1} & \qw \\
\lstick{\ket{b}} & \ctrl{1} & \qw \\
\lstick{\ket{c}} & \gate{1} & \qw &\push{\ket{(\neg a \vee \neg b)\cdot c\vee a\cdot b}}}
\end{array}$}}\caption{Forced gate Faults: (a) $2-$CN gate that correctly acts on a
$\ket{1}$ by changing the state to $\ket{0}$, (b) $2-$CN gate that
correctly acts on a $\ket{0}$ by changing the state to $\ket{1}$.
(In both of these cases, we consider binary
inputs.)}\label{cir:Forced gate Fault_model}
\end{figure}

Forced gate faults are non-unitary.  An example of a non-unitary operation is a gate that when activated
applies an ``\emph{amplitude dampening process}~\cite{NielsenC:00}''
to the target bit (a type of relaxation
process~\cite{Vandersypen:NMRBasics:02}). A test set complete for Quantum Fault Model~\ref{Requirement:gates act
on full basis} forces the gate to act on both a $\ket{0}$ and a
$\ket{1}$ to uniquely show that this considered fault is not present.  This can be seen
further by examining the Truth Table in Fig.~\ref{tab:Truth Table
for Forced Gate Faults}.  The Forced Gate Fault model is given in
Fig.~\ref{cir:Forced gate Fault_model}.
\begin{Requirement}\label{Requirement:gates act on full basis}
    (\textbf{Forced Gate Fault Model}) Each target must separately act on basis state inputs $\ket{0}$ and $\ket{1}$.~$\blacksquare$
\end{Requirement}
%
%A complete
%test set for the fault model presented in Fig.~\ref{cir:Forced gate
%Fault_model} forces each gate in a binary quantum network to act on
%both basis input states $\ket{0}$ and $\ket{1}$.

\begin{figure}[h]
\small{\centerline{ $\begin{tabular}{|c||c|c|c|}
  \hline
  \emph{input} & GC & a & b \\
  \hline
  $\ket{000}$ & 000 & 000 & 000 \\
    \hline
  $\ket{001}$ & 001 & 001 & 001 \\
    \hline
  $\ket{010}$ & 010 & 010 & 010 \\
    \hline
  $\ket{011}$ & 011 & 011 & 011 \\
    \hline
  $\ket{100}$ & 100 & 100 & 100 \\
    \hline
  $\ket{101}$ & 101 & 101 & 101 \\
    \hline
  $\ket{110}$ & 111 & 110 & 111 \\
    \hline
  $\ket{111}$ & 110 & 110 & 111 \\
  \hline
\end{tabular}$}}
\caption{The Truth Table for $2-$CN Gate and the impact of Forced
Gate Faults: The column denoted 'input' shows the respective input
combinations possible. Columns denoted 'a'
and 'b' show the circuit's response given the presence of faults
from Fig.~\ref{cir:Forced gate Fault_model} (a) and
(b), respectively.}\label{tab:Truth Table for Forced Gate Faults}
\end{figure}

\begin{figure}[h]
\small{\centerline{ $\begin{tabular}{|c||c|c|}
  \hline
  \emph{input} & a & b \\
  \hline
  $\ket{000}$ & 0 & 0 \\
    \hline
  $\ket{001}$ & 0 & 0 \\
    \hline
  $\ket{010}$ & 0 & 0 \\
    \hline
  $\ket{011}$ & 0 & 0 \\
    \hline
  $\ket{100}$ & 0 & 0 \\
    \hline
  $\ket{101}$ & 0 & 0 \\
    \hline
  $\ket{110}$ & 1 & 0 \\
    \hline
  $\ket{111}$ & 0 & 1 \\
  \hline
\end{tabular}$}}
\caption{Fault Table for $2-$CN Gate perturbed by the Forced Gate
Faults given in Fig.~\ref{cir:Forced gate Fault_model}. Each binary
entry in the fault table corresponds to a single test (row) and a
single fault (column). Tests are labeled $\ket{000}$ to $\ket{111}$
and faults are labeled 'a' and 'b', as shown in Fig.~\ref{cir:Forced
gate Fault_model} (a) and (b).  A '1' in the table corresponds to a
given test (row) detecting a given fault (column).}\label{tab:Fault
Table for Forced Gate Faults}
\end{figure}

\section{Measurement Faults}\label{sec:Measurement Faults}
Certain types of measurement faults can be caused from a "\emph{limitation in the
sensitivity of a measurement apparatus}~\cite{percisionmeasure}."
Measurement faults are often modelled by the Pauli
Fault Model. For example, to project (measure) the state of a photon one places a
slit in front of a photo detector.
Polarization states inline with the slit will be allowed to reach
the photo detector and the angle of the slit is subject to error~\cite{qubitMeasure:01}.
Aside from the Pauli Fault Model already considered, a faulty measurement instrument is modelled as a probe that
couples to a qubit and consistently returns an a certain value.  In Fig.~\ref{cir:Measurement Errors Impacting a $2-$CN Gate}
the single measurement fault model is illustrated by placing a
faulty measurement gate at the output of the circuit. The truth table
derived from Fig.~\ref{cir:Measurement Errors Impacting a $2-$CN
Gate} is shown in Fig.~\ref{tab:Truth Table Measurement Faults}. The
corresponding fault table if given in Fig.~\ref{tab:Fault Table
Measurement Faults}.  Test sets complete for the Measurement Fault Model~\ref{Requirement:Measurement} detect the the faults defined below in Def.~\ref{def:Measurement Single Fault Model}.

\begin{definition}\label{def:Measurement Single Fault Model}
    Measurement Fault Model: A working
    measurement gate is replaced with a faulty measurement gate that returns only a \emph{logic-zero} or a \emph{logic-one}.
\end{definition}
\begin{Requirement}\label{Requirement:Measurement}
    (\textbf{Measurement Fault Model}) Each qubit must be measured in both \emph{logic-zero} and
    \emph{logic-one} states.~$\blacksquare$
\end{Requirement}

\begin{figure}[h]\small{\centerline{ $\begin{array}{ccc}
\Qcircuit @C=.5em @R=.5em  {
  & \mbox{~(a) } & &&&&\\
\lstick{\ket{a}} & \ctrl{1} & \measuretab{0} \\
\lstick{\ket{b}} & \ctrl{1} & \meter \\
\lstick{\ket{c}} & \targ & \meter }
 & \Qcircuit @C=.5em @R=.5em  {
  & \mbox{~(b) } & &&&&\\
\lstick{\ket{a}} & \ctrl{1} & \meter \\
\lstick{\ket{b}} & \ctrl{1} & \measuretab{0}\\
\lstick{\ket{c}} & \targ    & \meter}
 & \Qcircuit @C=.5em @R=.5em  {
  & \mbox{~(c) } & &&&&\\
\lstick{\ket{a}} & \ctrl{1} & \meter \\
\lstick{\ket{b}} & \ctrl{1} & \meter \\
\lstick{\ket{c}} & \targ    & \measuretab{0}}
\\\\
\Qcircuit @C=.5em @R=.5em  {
  & \mbox{~(d) } & &&&&\\
\lstick{\ket{a}} & \ctrl{1} & \measuretab{1}\\
\lstick{\ket{b}} & \ctrl{1} & \meter\\
\lstick{\ket{c}} & \targ    & \meter}
 & \Qcircuit @C=.5em @R=.5em  {
  & \mbox{~(e) } & &&&&\\
\lstick{\ket{a}} & \ctrl{1} & \meter \\
\lstick{\ket{b}} & \ctrl{1} & \measuretab{1} \\
\lstick{\ket{c}} & \targ    & \meter}
 & \Qcircuit @C=.5em @R=.5em  {
  & \mbox{~(f) } & &&&&\\
\lstick{\ket{a}} & \ctrl{1} & \meter\\
\lstick{\ket{b}} & \ctrl{1} & \meter\\
\lstick{\ket{c}} & \targ    & \measuretab{1}}\\\\
\end{array}$}} \caption{Measurement Errors:
Figs (a), (b) and (c) illustrate measurement faults that
statistically favor \emph{logic-zero}. Figs (d), (e) and (f) contain
measurement faults statistically favoring
\emph{logic-one}.}\label{cir:Measurement Errors Impacting a $2-$CN
Gate}
\end{figure}

\begin{figure}[h]
\small{\centerline{
\begin{tabular}{|c||c|c|c|c|c|c|c|}
  \hline
  \emph{input} & GC & a & b & c & d & e & f \\
  \hline
  $\ket{000}$ & 000 & 000 & 000 & 000 & 100 & 010 & 001 \\
  \hline
  $\ket{001}$ & 001 & 001 & 001 & 000 & 101 & 011 & 001 \\
  \hline
  $\ket{010}$ & 010 & 010 & 000 & 010 & 110 & 010 & 011 \\
  \hline
  $\ket{011}$ & 011 & 011 & 001 & 010 & 111 & 011 & 011 \\
  \hline
  $\ket{100}$ & 100 & 000 & 100 & 100 & 100 & 110 & 101 \\
  \hline
  $\ket{101}$ & 101 & 001 & 101 & 100 & 101 & 111 & 101 \\
  \hline
  $\ket{110}$ & 111 & 010 & 100 & 110 & 110 & 110 & 111 \\
  \hline
  $\ket{111}$ & 110 & 011 & 101 & 110 & 111 & 111 & 111 \\
  \hline
\end{tabular}
} }\caption{Truth Table for $2-$CN Gate Impacted by Measurement
Faults. The column denoted 'input' shows the input combinations
possible on the amplitude plane. Columns denoted 'a' through 'f'
show the circuit's response given the presence of faults from
Fig.~\ref{cir:Measurement Errors Impacting a $2-$CN Gate} (a)
through (f).}\label{tab:Truth Table Measurement Faults}
\end{figure}

\begin{figure}[h]\small{
\centerline{
\begin{tabular}{|c||c|c|c|c|c|c|}
  \hline
  \emph{input} & a & b & c & d & e & f \\
  \hline
  $\ket{000}$ & 0 & 0 & 0 & 1 & 1 & 1 \\
    \hline
  $\ket{001}$ & 0 & 0 & 1 & 1 & 1 & 0 \\
    \hline
  $\ket{010}$ & 0 & 1 & 0 & 1 & 0 & 1 \\
    \hline
  $\ket{011}$ & 0 & 1 & 1 & 1 & 0 & 0 \\
    \hline
  $\ket{100}$ & 1 & 0 & 0 & 0 & 1 & 1 \\
    \hline
  $\ket{101}$ & 1 & 0 & 1 & 0 & 1 & 0 \\
    \hline
  $\ket{110}$ & 1 & 1 & 0 & 0 & 0 & 1 \\
    \hline
  $\ket{111}$ & 1 & 1 & 1 & 0 & 0 & 0 \\
  \hline
\end{tabular}}
} \caption{Fault table for $2-$CN gate derived from
Fig.~\ref{cir:Measurement Errors Impacting a $2-$CN Gate}. Each
binary entry in the fault table corresponds to a single test (row)
and a single fault (column). Tests are labeled $\ket{000}$ to
$\ket{111}$ and faults are labeled 'a' through 'f'.  A value of '1'
in the table corresponds to a given test (row) detecting (covering)
a given fault (column).}\label{tab:Fault Table Measurement Faults}
\end{figure}

\section{Conclusion}\label{sec:con}

In this work we have considered the adaptation of know classical methods to test quantum circuits.  The first step in this program is to develop a set of realistic quantum fault models.  The fault models we have presented enable circuit testing at the logical, as oposed to currently practiced approach of testing quantum circuits at the device level.  Finding additional fault models, that are physically inspired and not dominated by the those introduced here, is an open problem.  In the appendix we have included an example test set for a specfic circuit using the fault models presented in this work.  

%This work has justified several logical fault models for quantum circuits.  The fault models
%that have been justified are: Initialization Errors
%(Requirement~\ref{Requirement:initialization fault model},
%Sec.~\ref{subsec:Initialization Errors}), Measurement Faults
%(Requirement~\ref{Requirement:Measurement}, Sec.~\ref{subsec:Measurement
%Faults}), Lost Phase Faults
%(Requirements~\ref{Requirement:phase1}~and~\ref{Requirement:lost phase},
%Sec.~\ref{subsec:Phase Faults}), Faded Control Faults
%(Requirement~\ref{Requirement:fadedcontrol}, Sec.~\ref{subsec:Control Faults}),
%and Faded gate faults (Requirement~\ref{Requirement:gates act on full basis},
%Sec.~\ref{subsec:Broken Gate Faults}).

\section*{Acknowledgments}
The Quantum Circuit diagrams were drawn in \LaTeX \ using
Q-circuit~\cite{qcircuit}.  The simulation tool
QuIDDPro~\cite{ViamontesMH:05} was used during this study.
We would like to thank G.F. Viamontes, D. Maslov, J.A. Jones and P.J. Love.

\appendix
\section{The Partial Trace}\label{sec:The Partial Trace}
In this Appendix we review the partial trace~\cite{NielsenC:00}.
Consider Hilbert space $A$ of system $A\otimes B$. We
will trace over system $A$, leaving system $B$ in a
mixed state. We have, $tr_A(\ket{a_1}\bra{a_2}\otimes
\ket{b_1}\bra{b_2})$. Now consider ${\ket{e_i}}$ as an orthonormal
basis for system $A$, we may write the partial trace as, $\sum_i
\bra{e_i}\ket{a_1}\bra{a_2}\otimes \ket{b_1}\bra{b_2}\ket{e_i}$.
After we recall the general fact about tensor products,
$\ket{a}\bra{b}\otimes \ket{c}\bra{d}$ =
$\ket{a}\ket{c}\bra{b}\bra{d}$, it is easy to see a well known
equation for the trace of a component part of a composite system,
$\sum_i\bra{e_i}\ket{a_1}\bra{a_2}\ket{e_i}\otimes\ket{b_1}\bra{b_2}$
=$tr(\ket{a_1}\bra{a_2})\ket{b_1}\bra{b_2}$.

\section{Current Methods Used to Test Quantum Circuits}\label{sec:Current Methods Used to Test Quantum Circuits}

In this Appendix we review the main approach currently used to test
quantum circuits.  In the mid to late $90$'s experimentalists developed a method of
black box characterization known as quantum process
tomography~\cite{blackbox}. A quantum process is described as a map
between input and output quantum states, e.g. $\rho_{out} =
\mathcal{E}(\rho_{in})=\sum_j E_j\rho_{in} E_j^\dagger$, where the
map $\mathcal{E}$ is a \emph{quantum operation}\footnote{In this
work we only consider the case where $\sum_j E_j E_j^\dagger = I$.}
and the operators $E_j$ are called operation elements.\footnote{A
review of the properties of operation elements is given in Ch. 3 of
the 1998 PhD Thesis by Nielsen~\cite{Nielsen:QI:PhD:98}.} Process
tomography is a procedure used to reconstruct the
behavior of a quantum network by performing state tomography on a
set of initial states $\rho_i$ that form an operator basis for the
system in question~\cite{Childes}.\footnote{A purely mathematical
discussion of process tomography is presented, all measurements are
treated as yielding exact probabilities and all sources of error in
those measurements are ignored.  For experimental background see for
example~\cite{Obrien:QPT_CNOT:04}
and~\cite{DWLeungPhD:00, qubitMeasure:01}. Chapter $8$ in
Ref.~\cite{NielsenC:00} also has an introduction to both State and
Process Tomography.} The input states and measurement projectors in
process tomography each form a basis for the set of $n-$qubit
density matrices requiring $d^2 = 2^{2n}$ elements in each
set~\cite{NielsenC:00}, where $d$ is the dimension of the Hilbert
space. For a two-qubit gate $d^2 = 16$, resulting in $256$ different
settings of input states and measurement projectors. One of many
possible input combinations (adapted from the optics experiment
in~\cite{Obrien:QPT_CNOT:04}) forming an operator basis needed to
characterize the space of two-qubit circuits is the
following:\footnote{Using the notation that:
($\ket{y_+}=\ket{0}+i\ket{1}$ and $\ket{y_-}=\ket{0}-i\ket{1}$). The
measurement projectors corresponding to this set of initial states
adapted from the optics experiment given
in~\cite{Obrien:QPT_CNOT:04} are: $\bra{00}$, $\bra{10}$,
$\bra{+1}$, $\bra{y_-0}$, $\bra{y_-1}$, $\bra{11}$, $\bra{01}$,
$\bra{0-}$, $\bra{0y_-}$, $\bra{y_-y_-}$, $\bra{y_--}$, $\bra{+-}$,
$\bra{+y_+}$, $\bra{1-}$, $\bra{1y_-}$ and $\bra{+0}$. }
\begin{eqnarray}\label{eqn:supperposstate}
\{ \ket{00}, \ket{01}, \ket{10}, \ket{11}, \ket{0+}, \ket{0y_-},\ket{1y_-}, \ket{1+},\nonumber\\
 \ket{++}, \ket{y_+y_-}, \ket{y_++},\ket{+y_+}, \ket{+1}, \nonumber\\
 \ket{y_+1}, \ket{+0}, \ket{y_+0}\}.
\end{eqnarray}
Of course there exist many possible choices for such a basis. In
general however, for a system of $n$ qubits the computational basis
states $\ket{0},...,\ket{2^{n-1}}$ and superpositions
$(\ket{q}\pm\ket{r})/\sqrt{2}$ are prepared, where $q\neq
r$~\cite{averageGateFidelity:02}.

Given many copies of an experimental sample, state tomography is a
procedure allowing one to reconstruct an arbitrary quantum state to
a given accuracy.  It requires a set of simple measurement operators
that are products of Pauli matrices.
The method relies on creating a set
of orthogonal measurements and using the Hilbert-Schmidt inner
product~\cite{NielsenC:00} to expand the state of $\rho$ based on
the average outcome of each measurement.  A single qubit may be
reconstructed as the following density matrix:
\begin{equation}\label{eqn:expansion of a single qubit}
\rho = \frac{tr(\rho)\sigma_{i} + tr(\sigma_x \rho)\sigma_x +
tr(\sigma_y \rho)\sigma_y + tr(\sigma_z \rho)\sigma_z}{2}.
\end{equation}
Expressions like $tr(\sigma_x \rho)$ in Eqn.~\ref{eqn:expansion of a
single qubit} refer to an average measurement outcome where
$\sigma_x$ is an observable. A similar expansion to that of
Eqn.~\ref{eqn:expansion of a single qubit} applies to $n$--qubit
systems.  For example, reconstruction of any two-qubit operator
requires a total of $2^{2n}=16$ measurement observables:
\begin{eqnarray}\label{eqn:supperposstate}
\{\sigma_i \otimes \sigma_i , \sigma_i \otimes \sigma_x, \sigma_i
\otimes \sigma_y, \sigma_i \otimes \sigma_z, \sigma_x \otimes
\sigma_i , \sigma_x \otimes \sigma_x, \sigma_x \otimes \sigma_y,
\sigma_x \otimes \sigma_z, \sigma_y \otimes \sigma_i , \sigma_y
\otimes \sigma_x, \nonumber\\
\sigma_y \otimes \sigma_y, \sigma_y \otimes \sigma_z, \sigma_z
\otimes \sigma_i , \sigma_z \otimes \sigma_x, \sigma_z \otimes
\sigma_y, \sigma_z \otimes \sigma_z\}.
\end{eqnarray}

A difficulty associated with quantum process tomography is that in
experimental practice, the observables are not easily realized. A system with $d$ dimensions requires $16^d - 4^d$ independent
parameters to uniquely describe the process~\cite{Childes}, where
$d=2^n$. The useful method of quantum process tomography was
developed out of a need for black box characterization (for that
purpose its use appears unavoidable).  However, process tomography
works independently of the set of gates realized in the network and
their possible faults and when used as a method to test quantum
switching networks, it has a classical counterpart known as
\emph{brute-force} or complete \emph{functional-testing}. %For example,

Distance measures between quantum
states are now reviewed.  First we recall the well known
Fidelity measure between quantum states.

\begin{definition}\label{def:Fidelity}
    The \textbf{Fidelity} between density matrices $\rho$ and
    $\sigma$ is defined as:
    \begin{equation}\label{eqn:Fidelity}
        F(\rho,\sigma)\equiv tr\left(\sqrt{\sqrt{\rho}\sigma \sqrt{\rho}}\right)^2
    \end{equation}
    When $\rho = \ket{\psi}\bra{\psi}$ is a pure state the fidelity has an easy
    interpretation as the overlap between $\rho$ and $\sigma$, reducing
    to:
    \begin{equation*}
        F(\psi, \sigma) = \bra{\psi}\sigma\ket{\psi}.
    \end{equation*}
    Furthermore, the Fidelity evaluates to zero when two pure states being compared are orthogonal, it evaluates to one when two states being compared are identical,
    and is not a metric.\footnote{Two common ways of turning the Fidelity into a metric are the \emph{Bures
    metric}, $B(\rho,\sigma)\equiv\sqrt{2-2\sqrt{F(\rho,\sigma)}}$ and the \emph{angle}, $A(\rho,\sigma)\equiv\arccos\left(\sqrt{F(\rho,\sigma)}\right)$,
    a very comprehensive discussion of these details can be found elsewhere, e.g. Ref.~\cite{distancemeasure}.}
    For a discussion regarding an operational interpretation of the Fidelity for a mixed state
    see Reference~\cite{fidelityInterp}.
\end{definition}

A second common distance measure is the Trace Distance between quantum states.

\begin{definition}\label{def:Trace}
     The \textbf{Trace Distance} between density matrices $\rho$ and
    $\sigma$ is defined as:
    \begin{equation}\label{eqn:Trace}
        D(\rho,\sigma)\equiv \frac{1}{2}tr|\rho-\sigma |
    \end{equation}
    where $|Z|=\sqrt{Z^\dagger Z}$.  Since $0 \leq D \leq 1$ the trace distance
    is a genuine metric on quantum states~\cite{distancemeasure, NielsenC:00}
and thus has the following three properties: $($\emph{i}$)$
$D(\rho,\sigma)\geq 0$ with $D(\rho,\sigma) = 0$ \emph{iff} $\sigma
= \rho$, $($\emph{ii}$)$ Symmetry: $D(\rho,\sigma) =
D(\sigma,\rho)$, and $($\emph{iii}$)$ the Triangle Inequality:
$D(\mathcal{E}(\rho),\mathcal{G}(\rho))\leq
D(\mathcal{E}(\rho),\mathcal{F}(\rho))+D(\mathcal{F}(\rho),\mathcal{G}(\rho))$.
The Trace Distance represents the statistical distribution between quantum
    states with respect to measurement.  The Trace Distance has the
    property of \emph{contractivity},
    $D(\mathcal{E}(\rho),\mathcal{E}(\sigma))\leq D(\rho,\sigma)$
    whenever $\mathcal{E}$ is a trace-preserving quantum
    operation.  This just means that acting on arbitrary quantum
    states $\rho$ and $\sigma$ both with operation $\mathcal{E}$ will
    never increase how well one can distinguish these states with respect to measurements~\cite{distancemeasure, NielsenC:00}.
\end{definition}

The Trace Distance and Fidelity are complementary measures and
should be considered equally important when comparing two quantum
states~\cite{distancemeasure}. Distance measures may also be used to compare and contrast a
real process $\mathcal{F}$ and an ideal process $\mathcal{E}$, such
that $\Delta(\mathcal{F},\mathcal{E})$ defines an error metric on a
quantum process~\cite{distancemeasure}.

\begin{definition}\label{def:s-fidelity}
     The \textbf{S-Fidelity} between real quantum process
    $\mathcal{F}$ and ideal quantum process $\mathcal{E}$ is defined as:
    \begin{equation}\label{eqn:s-fidelity}
        \Delta_{min}^F(\mathcal{F},\mathcal{E})\equiv
        \min_{\ket{\psi}} \Delta(\mathcal{F}(\psi),\mathcal{E}(\psi))
    \end{equation}
    where the minimum is over all possible pure state inputs and $\Delta$ is a Fidelity measure on quantum states.
\end{definition}

\begin{definition}\label{def:s-distance}
    The \textbf{S-Distance} between real quantum process $\mathcal{F}$ and
    ideal quantum process $\mathcal{E}$ is defined as:
    \begin{equation}\label{eqn:s-distance}
        \Delta_{max}^D(\mathcal{F},\mathcal{E})\equiv \max_{\ket{\psi}}\Delta(\mathcal{F}(\psi),\mathcal{E}(\psi))
    \end{equation}
    where the maximum is over all possible pure state inputs and $\Delta$ is a Distance metric on quantum states.

\end{definition}

Instead of considering all pure states, it is helpful to restrict our
thinking to a set of inputs needed to form a complete operator basis
for the system in question. In this case, experimentally determining
the S-Distance/S-Fidelity amounts to performing state tomography on
this complete operator basis input set while keeping track of the
worst Trace Distance (\ref{eqn:Trace})/Fidelity (\ref{eqn:Fidelity})
between the reconstructed state and that of the ideal.  Ref.~\cite{distancemeasure}
stated that, \emph{"...the S-Distance and S-Fidelity are the two
best error measures, and should be used as the basis for comparison
of real quantum information processing experiments to the
theoretical ideal."}

%Due to the mentioned limitations of
%process tomography, recent literature has risen seeking methods to
%quickly judge whether a quantum circuit is within tolerance required
%for successful application, such as Bowdrey et.
%al.~\cite{fidelitySQM:02}, and later
%Nielsen~\cite{averageGateFidelity:02}. Of particular interest to our
%goal is the work by Gilchrist, Langford, and
%Nielsen~\cite{distancemeasure}.  This paper influenced our point of
%view by presenting a logical set of Requirements that an error measure
%$\Delta$ between the real process $\mathcal{F}(\psi)$ and ideal
%process $\mathcal{E}(\psi)$ should adhere to. They developed
%criteria allowing experimentalists to avoid full process tomography
%by applying "\emph{distance measures}" between the circuit's
%response to a set of input states forming a complete operator
%basis~\cite{distancemeasure}, as opposed to reconstructing the
%operator in full process tomography. This is done by comparing the
%real process $\mathcal{F}(\psi)$ with the ideal process
%$\mathcal{E}(\psi)$ while keeping track of the worst case Trace
%Distance (\emph{S-Distance}) and worst case Fidelity
%(\emph{S-Fidelity}) for all considered input states. In light of a
%"\emph{gold standard}\footnote{Given in Ref.~\cite{distancemeasure}
%is the notion of a "\emph{gold standard}" for quantum information
%processing. A "\emph{gold standard}" is a single measure of distance
%$\Delta$ that may be used to compare and contrast a real process
%$\mathcal{F}$ and an ideal process $\mathcal{E}$,
%$\Delta(\mathcal{F},\mathcal{E})$.}"

\section{Testing Example}\label{sec:example}

In this section we will present the following quantum circuit.  Figure~\ref{cir:testexample} shows diagramatically the quantum circuit perturbed by some of the fault models we have introduced here. The interested reader can consult Table~\ref{tab1} to consider the test set that will determine if any of the fault models considered in this work are present or not.
\begin{center}\small{\centerline{
\Qcircuit @C=.4em @R=.4em  {
                 &          & \mbox{GC} && \\
\lstick{\ket{a}} & \ctrl{1} & \ctrl{1} & \qw & \push{\ket{a}} \\
\lstick{\ket{b}} & \targ    & \ctrl{1} & \qw & \push{\ket{a\oplus b}} \\
\lstick{\ket{c}} & \qw      & \targ    & \qw & \push{\ket{a\cdot\neg b\oplus c}}}
}}
\end{center}
%\begin{figure}[h]\small{
%\centerline{

\begin{center}
\begin{tabular}{c|c|c|c|c|c|c|c|c}
  \hline
  abc & $GC$ & $f_1$ & $f_2$ & $f_3$ & $f_4$ & $f_5$ & $f_6$ & $f_7$ \\
  000 & 000  & 010   & 000   & 000   & 000   & 000   & 000   & 000 \\
  001 & 001  & 011   & 001   & 001   & 001   & 001   & 001   & 001 \\
  010 & 010  & 000   & 011   & 010   & 010   & 010   & 010   & 010 \\
  011 & 011  & 001   & 010   & 011   & 011   & 011   & 011   & 011 \\
  100 & 111  & 111   & 111   & 101   & 100   & 111   & 110   & 110 \\
  101 & 110  & 110   & 110   & 100   & 101   & 110   & 111   & 111 \\
  110 & 100  & 100   & 100   & 111   & 100   & 111   & 100   & 101 \\
  111 & 101  & 101   & 101   & 110   & 101   & 110   & 100   & 101 \\
  \hline
\end{tabular}\label{tab1}
\end{center}

\begin{figure}[h]\small{\centerline{
$\begin{array}{cccccccc}
  \Qcircuit @C=.4em @R=.7em  {
                 &          & \mbox{$f_1$} && \\
\lstick{\ket{a}} & \qw      & \ctrl{1} & \qw & \push{\ket{a}} \\
\lstick{\ket{b}} & \targ    & \ctrl{1} & \qw & \push{\ket{\neg b}} \\
\lstick{\ket{c}} & \qw      & \targ    & \qw & \push{\ket{a\cdot\neg b\oplus c}}\gategroup{2}{2}{2}{2}{.8em}{..} ~~~~~~~~~~~~~~}
 &  ~~~~~~~~~~~~~~\Qcircuit @C=.4em @R=.7em  {
                 &          & \mbox{$f_2$} && \\
\lstick{\ket{a}} & \ctrl{1} & \qw      & \qw & \push{\ket{a}} \\
\lstick{\ket{b}} & \targ    & \ctrl{1} & \qw & \push{\ket{a\oplus b}} \\
\lstick{\ket{c}} & \qw      & \targ    & \qw & \push{\ket{a\oplus b\otimes c}}\gategroup{2}{3}{2}{3}{.8em}{..}}~~~~~~
 &  ~~~~~~\Qcircuit @C=.4em @R=.7em  {
                 &          & \mbox{$f_3$} && \\
\lstick{\ket{a}} & \ctrl{1} & \ctrl{2} & \qw & \push{\ket{a}} \\
\lstick{\ket{b}} & \targ    & \qw      & \qw & \push{\ket{a\oplus b}} \\
\lstick{\ket{c}} & \qw      & \targ    & \qw & \push{\ket{a\cdot c}}\gategroup{3}{3}{3}{3}{.8em}{..}}~~~~~~ ~~~~~~ &
 ~~~~~~\Qcircuit @C=.4em @R=.7em  {
                 &          & \mbox{$f_4$} && \\
\lstick{\ket{a}} & \ctrl{1} & \ctrl{1} & \qw & \push{\ket{a}} \\
\lstick{\ket{b}} & \gate{0}    & \ctrl{1} & \qw & \push{\ket{\neg a\cdot b}} \\
\lstick{\ket{c}} & \qw      & \targ    & \qw & \push{\ket{c}}}  \\\\
  \Qcircuit @C=.4em @R=.7em  {
                 &          & \mbox{$f_5$} && \\
\lstick{\ket{a}} & \ctrl{1} & \ctrl{1} & \qw & \push{\ket{a}} \\
\lstick{\ket{b}} & \gate{1}    & \ctrl{1} & \qw & \push{\ket{a\vee \neg a\cdot b}} \\
\lstick{\ket{c}} & \qw      & \targ    & \qw & \push{\ket{a\cdot c}}}
 &
   \Qcircuit @C=.4em @R=.7em  {
                 &          & \mbox{$f_6$} && \\
\lstick{\ket{a}} & \ctrl{1} & \ctrl{1} & \qw & \push{\ket{a}} \\
\lstick{\ket{b}} & \targ    & \ctrl{1} & \qw & \push{\ket{a\oplus b}} \\
\lstick{\ket{c}} & \qw      & \gate{0}    & \qw & \push{\ket{(\neg a\vee \neg b)\cdot c}}}
 & \Qcircuit @C=.4em @R=.7em  {
                 &          & \mbox{$f_7$} && \\
\lstick{\ket{a}} & \ctrl{1} & \ctrl{1} & \qw & \push{\ket{a}} \\
\lstick{\ket{b}} & \targ    & \ctrl{1} & \qw & \push{\ket{a\oplus b}} \\
\lstick{\ket{c}} & \qw      & \gate{1}    & \qw & \push{\ket{c\vee a\cdot \neg b}}} &
  \Qcircuit @C=.4em @R=.7em  {
                 &          & \mbox{$f_8$/$f_9$} && \\
\lstick{\underline{\ket{0/1}}} & \ctrl{1} & \ctrl{1} & \qw & \push{\ket{0/1}} \\
\lstick{\ket{b}} & \targ    & \ctrl{1} & \qw & \push{\ket{a\oplus b}} \\
\lstick{\ket{c}} & \qw      & \targ    & \qw & \push{\ket{c/ b\oplus c}}}
 \\\\  \Qcircuit @C=.4em @R=.7em  {
                 &          & \mbox{$f_{10}$/$f_{11}$} && \\
\lstick{\ket{a}} & \ctrl{1} & \ctrl{1} & \qw & \push{\ket{a/\neg a}} \\
\lstick{\underline{\ket{0/1}}} & \targ    & \ctrl{1} & \qw & \push{\ket{a\oplus b}} \\
\lstick{\ket{c}} & \qw      & \targ    & \qw & \push{\ket{a\cdot c/\neg a\oplus c}}} &
 \Qcircuit @C=.4em @R=.7em  {
                 &          & \mbox{$f_{12}$/$f_{13}$} && \\
\lstick{\ket{a}} & \ctrl{1} & \ctrl{1} & \qw & \push{\ket{a}} \\
\lstick{\ket{b}} & \targ    & \ctrl{1} & \qw & \push{\ket{a\oplus b}} \\
\lstick{\underline{\ket{0/1}}} & \qw      & \targ    & \qw & \push{\ket{a\cdot b/a\cdot b\oplus 1}}}
  & \Qcircuit @C=.4em @R=.4em  {
                 &          & \mbox{$f_{14}$/$f_{15}$} && \\
\lstick{\ket{a}} & \ctrl{1} & \ctrl{1} & \qw & \measuretab{0/1} \\
\lstick{\ket{b}} & \targ    & \ctrl{1} & \qw & \meter \\
\lstick{\ket{c}} & \qw      & \targ    & \qw & \meter} &
 \Qcircuit @C=.4em @R=.4em  {
                 &          & \mbox{$f_{16}$/$f_{17}$} && \\
\lstick{\ket{a}} & \ctrl{1} & \ctrl{1} & \qw & \meter \\
\lstick{\ket{b}} & \targ    & \ctrl{1} & \qw & \measuretab{0/1} \\
\lstick{\ket{c}} & \qw      & \targ    & \qw & \meter}
\\\\ \Qcircuit @C=.4em @R=.4em  {
                 &          & \mbox{$f_{17}$/$f_{18}$} && \\
\lstick{\ket{a}} & \ctrl{1} & \ctrl{1} & \qw & \meter \\
\lstick{\ket{b}} & \targ    & \ctrl{1} & \qw & \meter \\
\lstick{\ket{c}} & \qw      & \targ    & \qw & \measuretab{0/1}}
& \Qcircuit @C=.3em @R=.4em  {
                 &              && \mbox{$f_{19}$} && \\
\lstick{\ket{a}} & \gate{H}     & \qswap       & \ctrl{1} & \qw & \gate{H}     & \meter \\
\lstick{\ket{b}} & \gate{H}     & \targ\qwx    & \ctrl{1} & \qw & \gate{H}     & \meter \\
\lstick{\ket{c}} & \gate{H}     & \qw          & \targ    & \qw & \gate{H}     & \meter}
& \Qcircuit @C=.3em @R=.4em  {
                 &&          & \mbox{$f_{20}$} && \\
\lstick{\ket{a}} &\gate{H}     & \ctrl{1} & \qswap       & \qw &\gate{H}     & \meter \\
\lstick{\ket{b}} &\gate{H}     & \targ    & \ctrl{1}\qwx & \qw &\gate{H}     & \meter \\
\lstick{\ket{c}} &\gate{H}     & \qw      & \targ        & \qw &\gate{H}     & \meter} &
 \Qcircuit @C=.3em @R=.4em  {
                 &&          & \mbox{$f_{21}$} && \\
\lstick{\ket{a}} &\gate{H}     & \ctrl{1} & \ctrl{1}   & \qw &\gate{H}     & \meter \\
\lstick{\ket{b}} &\gate{H}     & \targ    & \qswap     & \qw &\gate{H}     & \meter \\
\lstick{\ket{c}} &\gate{H}     & \qw      & \targ\qwx  & \qw &\gate{H}     & \meter}
\end{array}$}} \caption{Measurement Errors:
Figs (a), (b) and (c) illustrate measurement faults that
statistically favor \emph{logic-zero}. Figs (d), (e) and (f) contain
measurement faults statistically favoring
\emph{logic-one}.}\label{cir:testexample}
\end{figure}

\end{document}